\documentclass[prb,twocolumn,showpacs,superscriptaddress]{revtex4}
\usepackage{amsmath}
\usepackage{citesort}
\usepackage{graphicx}

\begin{document}
\title{Rung--singlet charge--ordering in $\alpha'$-NaV$_2$O$_5$}
\author{R.T. Clay} 
\affiliation{Department of Physics and Astronomy,
Mississippi State University, Mississippi State, MS 39762-5167} 
\affiliation{ERC Center for Computational Science, 
Mississippi State University, Mississippi State, MS 39762-9627}
\author{S. Mazumdar}
\affiliation{Department of Physics, University of Arizona, Tucson, AZ
85721} \date{\today}
\begin{abstract}
We show that the proposed zig--zag charge--ordering model for
$\alpha'$--NaV$_2$O$_5$ is incompatible with a single transition that
induces both charge--ordering and spin gap. We introduce a two--band
model for $\alpha'$--NaV$_2$O$_5$ within which the simultaneous
charge--spin order transition is driven principally by electron-phonon
interactions. The spin gap is due to the formation of local dimers,
thus explaining the weak magnetic effect on this transition.
\end{abstract}
 \maketitle
$\alpha'$-NaV$_2$O$_5$ has attracted considerable attention because of
a peculiar insulator--insulator transition at T$_c$ = 34 K with the
{\it simultaneous} appearance of charge--ordering (CO) and spin gap
(SG) \cite{Isobe96a}.  The material consists of V--O ladders coupled
through direct V--V bonds, with the V--O layers separated by layers of
Na$^+$ ions.  The average number of d-electrons per V-ion is 1/2.  The
insulating behavior above T$_c$ is understood if each V--O--V rung
containing one d--electron is considered a ``site'' within a
Mott--Hubbard picture.  The SG transition to a nonmagnetic insulator
\cite{Isobe96a} was originally explained within a ``chain model''
\cite{Carpy75a}, which assumes CO into alternate V$^{4+}$ and V$^{5+}$
chains at T $>$ T$_c$, and spin--Peierls (SP) transition in the
V$^{4+}$ chains at T $<$ T$_c$.  NMR demonstrations of only one kind
of V--ions \cite{Ohama99a} at T $>$ T$_c$ precludes this model.
Inequivalent V--ions are, however, found for T $\leq$ T$_c$.

\begin{figure}
\centerline{\resizebox{3.0in}{!}{\includegraphics{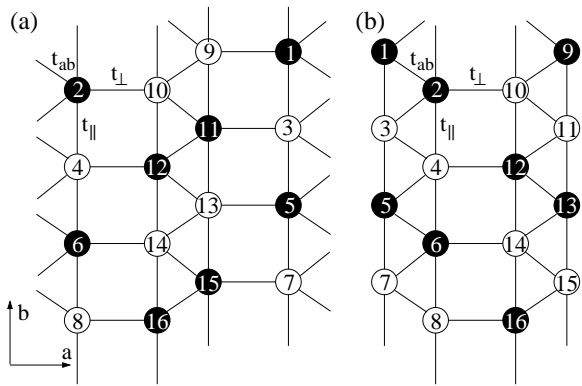}}}
\caption{Schematics of the proposed zig-zag CO. Black and white
circles correspond to V$^{4+}$ and V$^{5+}$, respectively.  (a) Two
ladders, periodic boundary conditions along $\hat{a}$ and $\hat{b}$.
(b) Single ladder with two neighboring chains, open boundary
conditions along $\hat{a}$.}
\label{fig1}
\end{figure}

Subsequently, there have been a number of theoretical discussions of
the low T state of $\alpha^\prime$-NaV$_2$O$_5$
\cite{Seo98a,Horsch98a,Chatterji98a,Gros99a,Mostovoy00a,Riera99a,Thalmeier98a},
also based on the one--band V--only picture.  Several of these recent
theories predict the so-called ``zig-zag'' CO (see Fig.~\ref{fig1})
\cite{Seo98a,Mostovoy00a}, which is driven by the Coulomb repulsion
between electrons occupying neighboring V ions within an extended
Hubbard model,
\begin{eqnarray}
H&=& -\sum_{<ij>,\sigma}t_{ij}(d^\dagger_{i\sigma}d_{j\sigma}+
d^\dagger_{j\sigma}d_{i\sigma}) \\ &+& U_0 \sum_i n^d_{i\uparrow}
n^d_{i\downarrow}+ U_1\sum_{<ij>} n^d_i n^d_j \nonumber
\end{eqnarray}
In the above, $d^\dagger_{i\sigma}$ creates an electron with spin
$\sigma$ on a V d--orbital,
$n^d_{i\sigma}=d^\dagger_{i\sigma}d_{i\sigma}$,
$n^d_i$=$n^d_{i\uparrow}+n^d_{i\downarrow}$, and $<ij>$ implies
nearest-neighbor V-ions. The hopping integrals $t_{ij}$ will be
denoted by $t_{\perp}$, $t_{||}$ and $t_{ab}$ for rung, leg and
interladder V--V bonds, respectively. For $U_0, U_1 >> |t_{ij}|$, the
ground state is the zig-zag CO of Fig.~\ref{fig1}. The origin of
the SG is not clear, and different theories have proposed different
bond distortions that generate the SG
\cite{Seo98a,Mostovoy00a,Riera99a}.

Experimentally, X-ray diffraction studies agree that below T$_c$ there
occurs a superlattice with 2$a$ $\times$ 2$b$ $\times$ 4$c$
supercell. The period doublings along $a$ and $b$ directions and the
quadruplings along the $c$ direction must originate from lattice
displacements. Anomalous X--ray scattering experiments have confirmed
CO, and have also claimed to have detected the zig--zag CO
\cite{Nakao00a,Grenier02a}, although other X-ray analysis have
suggested that the ordering occurs only on alternate ladders
\cite{Ludecke99a}.  Importantly, the relationship between the SG and
the CO is not obvious from these studies, and there is no consensus on
the pattern of bond distortions leading to the SG. Many experimental
investigations have focused on proving that the zig--zag CO is more
appropriate than the chain model, an approach that misses the
possibility of a third different CO. Optical measurements indicate a
strong role of phonons in the CO--SG transition
\cite{Popova02a}. Magnetic measurements have shown that (a) the spin
gap 2$\Delta_s$ is nearly twice that predicted from the BCS relation
2$\Delta_s$ = 3.5T$_c$ \cite{Fujii97a}, and (b) the effect of magnetic
fields on T$_c$ is much weaker than that expected for a simple SP
transition \cite{Fertey98a}.

\begin{figure}
\centerline{\resizebox{3.0in}{!}{\includegraphics{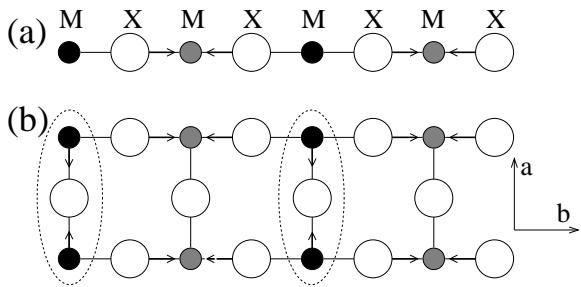}}}
\caption{Two-band models for (a) MX chains, (b) single ladder of
$\alpha'$--NaV$_2$O$_5$.  In (b) the large (small) circles correspond
to O(V)--ions.  Occupancy by electrons is larger (smaller) in the
black (gray) smaller circles in both (a) and (b). Singlet formation
occurs on the electron--rich rungs in (b).}
\label{fig2}
\end{figure}

There are two objectives of the present Letter. First, combining
recent theoretical results for 1/4--filled band systems
\cite{Clay03a,Vojta01a} and exact diagonalization studies we have
performed for finite two--dimensional (2D) clusters we show that
Eq.~(1) necessarily predicts {\it distinct} CO and SG transitions.
Second, we present a {\it two--band} model for $\alpha'$--NaV$_2$O$_5$
within which a single co-operative CO--SG transition is explained
naturally.  Specifically, we show that within the two--band model
incorporating both electron--electron (e--e) and electron--phonon
(e--ph) interactions, V--O--V ladder rungs are alternately
electron--rich and electron--poor (see Fig.~\ref{fig2}).  The SG is a direct
consequence of the CO, and is due to the formation of {\it local
singlets} on the electron--rich rung V--O--V bonds.  We point out that
there is no contradiction between this picture and the Mott--Hubbard
description of the T $>$ T$_c$ state.  Local dimers, emphasized in
several recent theories of correlated electrons
\cite{Clay03a,Sachdev03a}, were also discussed in earlier bipolaron theories
of Ti and V--oxides \cite{Chakraverty78a} and provide a new paradigm for
SG in 2D systems.

The fundamental problem with Eq.~(1) is that CO driven by the {\it
spin--independent} interaction $U_1$ should have occurred at higher T
where the free energy is dominated by high--spin excitations, and the
SG transition should have occurred {\it within the CO phase} at lower
T. This has also been recognized by other investigators
\cite{Thalmeier98a,Riera99a}.  Recent numerical work \cite{Vojta01a}
for the single ladder has shown that for the zig--zag CO to occur,
$U_1$ must be greater than a critical value $U_{1c} \geq 2|t_{||}|$,
as in one dimension (1D).  The magnitude of $U_{1c}$ {\it increases}
as $U_0$ decreases from infinity \cite{Vojta01a}, exactly as in 1D
\cite{Clay03a}. This result, taken together with the known result that
the ferromagnetic state with total spin S = S$_{max}$ is described by
the $U_0 \to \infty$ limit of Eq.~(1), implies that $U_{1c}(S =
S_{max}) < U_{1c}(S = 0)$.  Hence if $U_1 > U_{1c}$ in
$\alpha'--$NaV$_2$O$_5$, CO should occur at temperatures comparable to
the energy of the ferromagnetic state, while the SG would occur only
at lower T where singlet formation takes place.

An alternate scenario, not considered previously, is that $U_1 <
U_{1c}$ in $\alpha'$--NaV$_2$O$_5$ and the CO--SG transition is driven
by a co-operative lattice distortion. The SP transition in 1D
1/4--filled band systems with $U_1 < U_{1c}$, for example, is
accompanied by a CO, and has been referred to as a bond--charge
density wave \cite{Clay03a}.  There are two ways lattice--driven SG
transition and zig--zag CO can occur in $\alpha'$--NaV$_2$O$_5$: (i)
SP transitions involving the zig--zag bonds of single ladders, or (ii)
bond distortions that involve the entire multiple--ladder system. We
postpone the discussion of (i) until after we have presented the
two--band model, when we discuss why this transition is not possible.
The co-operative nature of the zig--zag CO and bond distortions in the
multiple--ladder systems has been discussed before \cite{Mostovoy00a},
albeit from a different perspective.  First, the difference in the
site occupancies can drive period 4 bond distortions among the
interladder bonds \cite{Seo98a,Riera99a}. Next, consider nearest
neighbor electron hops along the legs of the ladders in Fig.~\ref{fig1}. In
addition to the direct hops, there occur ``indirect hops'' between
them that involve a V--ion of the neighboring ladder. Since the
indirect hops involve either a charge--rich ion or a charge--poor ion
of the neighboring ladder, consecutive leg bonds become inequivalent
upon CO formation. Similarly, inequivalent leg bonds can drive a
dimerization of the intraladder zig--zag bonds in the lattice of
Fig.~\ref{fig1}(b) \cite{Mostovoy00a}. {\it Note that the coupled CO--bond
distortion necessarily implies that these arguments can be reversed,
and bond distortions can be considered to drive the CO \cite{Clay03a}.}

The tendency to bond distortions is measured within electronic models
from calculations of the bond orders \cite{Clay03a}, defined as,
\begin{equation}
B_{i,j}=\sum_{\sigma} \langle d^\dagger_{i\sigma} d_{j\sigma}
+d^\dagger_{j\sigma} d_{i\sigma}\rangle.
\label{bij}
\end{equation}
\begin{table}
\caption{\label{table1}Exact bond orders for the 16-site lattices in
Fig.~\ref{fig1}. All entries are for lattice of Fig.~\ref{fig1}(a) except
the zig-zag bonds labeled with $^*$, which are for lattice
of Fig.~\ref{fig1}(b). }
\begin{ruledtabular}
\begin{tabular}{|c||c|c|c|c|} 
&\multicolumn{2}{c|}{S$_z$=0} &\multicolumn{2}{c|}
{S$_z$=4} \\ \hline
Bond &\parbox[c][0.3in][c]{0.65in}{$t_{ab}=.05$eV $\Delta$n=0.238}&
\parbox{0.65in}{$t_{ab}=.1$eV $\Delta$n=0.191}&
\parbox{0.65in}{$t_{ab}=.05$eV $\Delta$n=0.391}&
\parbox{0.65in}{$t_{ab}=.1$eV $\Delta$n=0.368}\\ \hline \hline
$B_{1,2}$ &0.0280 &0.1425 &0.0169 &0.0188 \\
$B_{2,3}$ &0.0275 &0.1244 &0.0472 &0.0869 \\
$B_{3,4}$ &0.0261 &0.1187 &0.0168 &0.0178 \\
$B_{4,5}$ &0.0275 &0.1244 &0.0471 &0.0869 \\\hline\hline
$B_{2,4}$ &0.1795 &0.1893 &0.0807 &0.1054 \\
$B_{4,6}$ &0.1800 &0.1958 &0.0780 &0.0934 \\\hline \hline
$B_{2,12}^*$&0.1731 &0.1916 &0.0917 &0.0921 \\
$B_{12,6}^*$&0.1546 &0.0251 &0.0849 &0.0632 \\
\end{tabular}
\end{ruledtabular}
\end{table}
Differences in the bond orders corresponding to equivalent bonds
indicate {\it spontaneous} distortions that would occur for 0$^+$
e--ph coupling \cite{Clay03a}.  In Fig.~\ref{fig1} we show the finite
lattices for which we have done exact diagonalization studies.  CO is
achieved in our finite cluster calculations by adding a site--energy
term $\sum_i \epsilon^d_i n^d_i$ to Eq.~(1), where $\epsilon^d_i$ is
negative (positive) for the charge--rich (charge--poor) V--ions of
Fig.~\ref{fig1}, and can originate from e--ph coupling
\cite{Riera99a}.  Our calculations are for \cite{Smolinski98a} $U_0$ =
1 -- 4 eV, $t_{\perp}$ = 0.35 eV, $t_{||}$ = 0.15 eV and $t_{ab}$ =
0.05 -- 0.2 eV, and $0 \leq U_1 \leq$ 0.5 eV.  Since the CO and bond
order differences are driven by the site energies in our calculations,
our results are qualitatively the same for both $U_1$ = 0 and $U_1
\neq$ 0, and we report the results for $U_1$ = 0 only. In
Table~\ref{table1} we present the results for $U_0$ = 3 eV and
$|\epsilon^d_i|$ = 0.1 eV for two values of $t_{ab}$.  For both values
of $t_{ab}$ there occur period 4 bond distortions along the
interladder bonds and period 2 distortions of the ladder leg bonds in
the S = 0 state. In addition there occur period 2 bond distortions
along the zig--zag bonds for the lattice of Fig.~\ref{fig1}(b). As
mentioned above and discussed extensively in reference
\onlinecite{Clay03a}, it is not necessary that the CO drives the
lattice distortion. Rather, the bond distortions of Table~\ref{table1}
can also drive a zig--zag CO.

To investigate one versus two transitions, we have studied total spin
S = S$_{max}$ = 4 and these results are also reported in
Table~\ref{table1}.  It is seen that the tendency to bond distortion
is probably even stronger in S = S$_{max}$ than in S = 0. Note in
particular that the difference in the charge densities between the
charge--rich and the charge--poor sites, $\Delta$n, is {\it larger} in
the S = S$_{max}$ state.  We have performed similar calculations also
for the case of CO on alternate ladders \cite{Ludecke99a}.  Once
again, $\Delta$n and the bond distortions are larger in the
ferromagnetic state than in the singlet state.  There is therefore no
reason for the coupled CO--bond distortion to occur only in the S = 0
state. We therefore conclude that even with nonzero e--ph coupling the
zig--zag CO does not explain the observed CO--SG transition in
$\alpha'$--NaV$_2$O$_5$. This is understandable: {\it exactly as there
are two chains per ladder, there are two possible zig--zag CO
patterns, and there is no fundamental difference between the chain and
zig--zag models}, except that a short range $U_1$ instead of long
range Madelung interaction has been invoked to stabilize one zig--zag
CO over the other \cite{Seo98a,Mostovoy00a}.

We now show that a different CO pattern emerges when we
include the O$^{2-}$ ions. The most natural two--band model
for $\alpha'$--NaV$_2$O$_5$ is the 2D extension of the 
Hamiltonian that has been successfully used to describe 1D
transition metal--halogen (M--X) chains \cite{Weber92a},
\begin{eqnarray}
H&=&\sum_{<ij>, \mu, \sigma}[-t_{1,2} +
\alpha\Delta_{j\mu,i\mu}](d^\dagger_{i\mu\sigma}p_{j\mu\sigma} +h.c.)
\\ &+&\sum_{<jij'>,\mu, \sigma}[\epsilon_{i\mu} -
\beta(u_{j'\mu} - u_{j\mu})]n_{i\mu} + 
\frac{K}{2}\sum_{<ij>,\mu}\Delta_{j\mu,i\mu}^2 \nonumber \\
&-& t_{ab}\sum_{<\mu\mu'>,<ij>,\sigma}
(d^\dagger_{i\mu\sigma}d_{j\mu'\sigma} + h.c.) + U_0\sum_i
n^d_{i\uparrow}n^d_{i\downarrow} \nonumber
\end{eqnarray}
In the above, $i$, $j$ can be both V and O; $d^\dagger_{i\mu\sigma}$
and $p^\dagger_{i\mu\sigma}$ create electrons in V d--orbitals and O
p--orbitals, respectively; $\mu$ is the ladder index; $\langle jij'
\rangle$ implies consecutive sites in any direction, and $\langle
\mu\mu' \rangle$ implies neighboring ladders.  The V--O hopping
integrals $t_1$ and $t_2$ are along the ladder legs and rungs,
respectively, and $\epsilon_i$ are the site energies ($\epsilon_V = +
\epsilon$, $\epsilon_O = - \epsilon$); $u_{i\mu}$ are the
displacements of the V or O ions from their equilibrium positions and
$\Delta_{j\mu,i\mu}=u_{i\mu} -u_{j\mu}$, $\alpha$ and $\beta$ are the
intersite and on-site e--ph coupling constants, and K is the spring
constant.  For simplicity we have chosen $\alpha$, $\beta$ and K to be
same for leg and rung bonds. The number operator $n_{i\mu}$
corresponds to both V and O. We have included the on--site Coulomb
interaction only for V. Because of the large negative site energy for
O, the Hubbard interaction for O--ions makes little difference in the
final results.  We have not explicitly included intersite Coulomb
interactions, the effects of which we discuss later.

We have calculated self--consistent solutions to the $U_0$ = 0 limit
of Eq.~(2) for both single ladders and coupled multiple--ladder
systems, with periodic boundary conditions along $a$ and $b$
directions for systems up to 320 atoms.  Typical parameters were $t_1$
= 0.9 eV, $t_2$ = 1.3 eV, and $\epsilon_V$ ($\epsilon_O$) = 2.5 eV (--
2.5 eV), which were derived from reference \onlinecite{Smolinski98a}.
\begin{figure}
\centerline{\resizebox{2.75in}{!}{\includegraphics{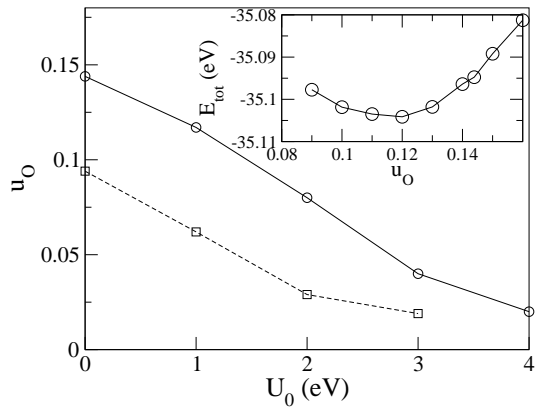}}}
\caption{The amplitude of the ladder-leg O--ion distortion versus the Hubbard 
repulsion for V--ions for electron--phonon coupling constants
$\lambda_{\alpha} = \alpha^2/Kt_1 = 0.44$, 
$\lambda_{\beta} = \beta^2/Kt_1 = 0.11$ (solid curve) and
$\lambda_{\alpha} = 0.28$ 
$\lambda_{\beta} = 0.07$ (dashed curve). Inset shows the
behavior for $U_0$ = 1 eV}
\label{fig3}
\end{figure}
Stable solutions for $\beta = 0^+$ correspond to breathing mode
distortions of the O--ions shown in Fig.~2(b) for the single ladder,
leading to (i) inequivalent charges on the V--ions on consecutive
ladder rungs, (ii) shorter V--O--V bonds on the rungs containing
nominally V$^{4+}$ ions, and (iii) period doubling along the $b$
direction. As there are two possible phases between adjacent ladders,
period doubling in the $a$ direction is possible as well
\cite{Riera99a}.  As indicated in Fig.~2(a), there is a one--to--one
correspondence between the distortions in the V--O ladders and the
M--X chains \cite{Weber92a}.  Each ladder rung is analogous to a
single M--site, with internal structure that leads to on--rung bond
distortion.  There also emerges a strong reason to ignore the SP
transition involving the zig--zag bonds within the single ladder (see
above). The distortion in Fig.~2(b) corresponds to alternating leg
O--O distances and inequivalent V charges; the SP distortion of the
intraladder zig--zag V--bonds requires alternating leg V--V distances
and nominally O$^{1-}$ and O$^{3-}$ anions. The impossibility of
obtaining O$^{3-}$ prevents this particular distortion. This result is
identical to the observation that M--M distances are uniform within
the two--band model for M--X chains \cite{Weber92a}.

The above two--band $U_0$ = 0 calculation is significant, as the
shorter bonds on the rungs containing the charge--rich V--ions suggest
singlet formation. CO and SG therefore occur simultaneously within
this model. The $U_0$ = 0 model is, however, unsuitable for the
insulating state observed at T $>$ T$_c$. A correct description of
$\alpha'$--NaV$_2$O$_5$ then requires nonzero but small $U_0$ within
Eq.~(3).  Specifically, $U_0$ should be large enough that the
Mott--Hubbard picture for insulating behavior for T $>$ T$_c$ is
qualitatively valid, and small enough that the rung--based CO is not
precluded. Note that the repulsion between two electrons on the same
rung bond is an {\it effective} Hubbard interaction $U_{eff}$ that is
much smaller than the bare $U_0$, while for small
$t_{ab}$, the effectively 1D 1/2--filled band ladder is insulating for
all $U_{eff}$.  For our model to be valid there must exist a range of
$U_0$ over which the distortion of Fig.~\ref{fig2} is nonvanishing and its
amplitude decreases slowly.

We have calculated the O--ion distortion $|u_O|$ for nonzero $U_0$
within Eq.~(3) for the single ladder with 4 rungs, using the
constrained path quantum Monte Carlo technique \cite{Zhang95a}.  We
retain the leg O--ions, but consider only effective V--V bonds on the
rungs, as the rung oxygens play a weak role in the CO and influence
only the intrarung distortion.  For different $\alpha$, $\beta$ and
$K$, we calculate the distortion $|u_O|$ at which the energy minimum
occurs for each $U_0$ (see insert, Fig.~\ref{fig3}). The overall
results for two different sets of e--ph coupling constants are shown
in Fig.~\ref{fig3}.  As expected, $|u_O|$ depends strongly on $\alpha$
and $\beta$, and also decreases with $U_0$. Significantly though,
while the decrease in $|u_O|$ is moderately rapid at small $U_0$, the
decrease is rather slow for $U_0 \geq$ 2 eV.  This behavior is a
signature of the persistence of the rung--based CO for realistic $U_0
\simeq 3 - 4$ eV.

A strong test of the rung--based CO model involves demonstrating the
co-operative nature of the CO--SG transition.  The calculated
$\Delta$n and bond distortions in Table~\ref{table1} for the zig--zag
CO are larger for S = S$_{max}$ than for S = 0. We have done the same
one--band calculation also for the rung distortion, for the lattice of
Fig.~1(b), and for isolated 6 and 8 rung ladders. The model
Hamiltonian is again the Hubbard Hamiltonian (Eq. ~1 with $U_1$ = 0)
including the site--energy term $\sum_i \epsilon^d_i n^d_i$, but now
the $\epsilon^d_i$ are equal for V--ions on the same rung, and
alternate in sign for consecutive rungs.  The results are identical in
all cases, viz., (i) bond orders are larger on rungs with larger
charge density in the S = 0 state, and (ii) the CO and bond distortion
is absent in S = S$_{max}$.  For the same $U_0$ and $\epsilon^d_i$ as
in Table~\ref{table1}, for example, $\Delta$n $\simeq$ 0.1 and $\Delta
B_{ij}$ $\simeq$ 0.05 in the S = 0 state of the single ladder with 8
rungs, while both $\Delta$n and $\Delta B_{ij}$ are $\simeq$ 0 in the
ferromagnetic state.  {\it In contrast to the zig--zag model
therefore, there occurs a single insulator--insulator transition
involving both charge and spin within the rung distortion picture.}
While the zig-zag CO could be obtained in a two-band model with oxygen
breathing distortions out of phase on the ladder legs
\cite{Mostovoy00a}, this would still give distinct CO and SG
transitions.

We now discuss the role of intersite Coulomb interactions, which were
ignored in Eq.~3. Intersite V--O Coulomb interactions have a weak
effect on the CO--SG transition, as might be expected.  We have
already argued that the V--V Coulomb interaction is smaller than the
critical value necessary to promote the zig--zag CO.  On the other
hand, the second neighbor V--V interaction, ignored in Eq.~1, strongly
destabilizes the zig--zag CO and promotes the rung CO. There exists
therefore a wide range of parameters for which the rung--based CO
dominates over the zig--zag CO.

In summary, we have shown that the zig--zag CO is not compatible with
a single CO--SG transition in $\alpha'$--NaV$_2$O$_5$, and have
proposed a two--band model for this system within which a co-operative
transition involving both charge and spin is driven by breathing mode
vibrations.  Strong d-p hybridization exceeding the energy scale of
$U_0$ in V-oxides has been suggested by Zimmermann et al., based on
photoemission studies \cite{Zimmermann98a}.  Interestingly, the
bond-order structure proposed in ref. \onlinecite{Chatterji98a} (see
Fig.~4) is consistent with rung ordering.  Our theory is consistent
with the observed strong phonon effects \cite{Popova02a}, while local
tightly bound singlet dimers would be in agreement with the observed
weak effect of magnetic field on the CO--SG transition
\cite{Fertey98a}. Careful reexaminations of the X-ray data are called
for. As mentioned already, many of these experiments were carried out
to distinguish between the chain and zig--zag CO's, which are the only
CO patterns possible within the one--band model. Our results
demonstrate that the tendency to form local dimers is particularly
strong at or near 1/4--filling of the band
\cite{Clay03a,Chakraverty78a}. Local dimer formation provides a new
mechanism for SG transitions beyond the more common mechanisms
involving SP dimerizations in 1D and spin frustrations, and it has
even been suggested that local dimers may be relevant in the context
of superconductivity in correlated electron systems
\cite{Clay03a,Sachdev03a}. It is highly interesting in this context
that (a) in superconducting $\beta$--Na$_{0.33}$V$_2$O$_5$ there occur
crystallographically three different kinds of V--ions, with perhaps
one class of V--band 1/4--filled \cite{Yamuchi02a}, and (b) the same
bandfilling characterizes organic superconductors.

We thank J.~T. Gammel, M.~V. Mostovoy, J.~L. Musfeldt and A. Painelli
for useful discussions. Work at Arizona was partially supported by the NSF.

\end{document}